\documentclass[aps,prl,bibnotes,reprint,twocolumn,showpacs,preprintnumbers,amsmath,amssymb,superscriptaddress,floatfix]{revtex4-2}
\usepackage[T1]{fontenc}
\usepackage[ansinew]{inputenc}
\usepackage{graphics}
\usepackage{dcolumn}
\usepackage{bm}
\usepackage{amsthm}
\usepackage{amsmath}
\usepackage{amssymb}
\usepackage{diagbox}
\usepackage{hyperref}
\usepackage{url}
\usepackage{xcolor}

\begin{document}

\title{Spin-disorder-induced angular anisotropy in polarized magnetic neutron scattering}

\author{Ivan Titov}
\affiliation{Department of Physics and Materials Science, University of Luxembourg, 162A~Avenue de la Faiencerie, 1511~Luxembourg, Grand Duchy of Luxembourg}

\author{Mathias Bersweiler}
\affiliation{Department of Physics and Materials Science, University of Luxembourg, 162A~Avenue de la Faiencerie, 1511~Luxembourg, Grand Duchy of Luxembourg}

\author{Michael P.\ Adams}
\affiliation{Department of Physics and Materials Science, University of Luxembourg, 162A~Avenue de la Faiencerie, 1511~Luxembourg, Grand Duchy of Luxembourg}

\author{Evelyn Pratami Sinaga}
\altaffiliation[Now at: ]{Department of Physics, Matana University, Gading Serpong, Tangerang, Banten~15810, Indonesia}\affiliation{Department of Physics and Materials Science, University of Luxembourg, 162A~Avenue de la Faiencerie, 1511~Luxembourg, Grand Duchy of Luxembourg}

\author{Venus Rai}
\affiliation{Department of Physics and Materials Science, University of Luxembourg, 162A~Avenue de la Faiencerie, 1511~Luxembourg, Grand Duchy of Luxembourg}

\author{\v{S}tefan Li\v{s}\v{c}\'{a}k}
\affiliation{Department of Physics and Materials Science, University of Luxembourg, 162A~Avenue de la Faiencerie, 1511~Luxembourg, Grand Duchy of Luxembourg}

\author{Max Lahr}
\affiliation{Department of Physics and Materials Science, University of Luxembourg, 162A~Avenue de la Faiencerie, 1511~Luxembourg, Grand Duchy of Luxembourg}

\author{Thomas L.\ Schmidt}
\affiliation{Department of Physics and Materials Science, University of Luxembourg, 162A~Avenue de la Faiencerie, 1511~Luxembourg, Grand Duchy of Luxembourg}

\author{Vladyslav M.\ Kuchkin}
\affiliation{Department of Physics and Materials Science, University of Luxembourg, 162A~Avenue de la Faiencerie, 1511~Luxembourg, Grand Duchy of Luxembourg}

\author{Andreas Haller}
\affiliation{Department of Physics and Materials Science, University of Luxembourg, 162A~Avenue de la Faiencerie, 1511~Luxembourg, Grand Duchy of Luxembourg}

\author{Kiyonori Suzuki}
\affiliation{Department of Materials Science and Engineering, Monash University, Clayton,
Victoria~3800, Australia}

\author{Nina-Juliane Steinke}
\affiliation{Institute Laue-Langevin, 71~Avenue des Martyrs, 38042~Grenoble, France}

\author{Diego Alba Venero}
\affiliation{ISIS Neutron and Muon Facility, Rutherford Appleton Laboratory, Didcot, OX110QX, United Kingdom}

\author{Dirk Honecker}
\affiliation{ISIS Neutron and Muon Facility, Rutherford Appleton Laboratory, Didcot, OX110QX, United Kingdom}

\author{Joachim Kohlbrecher}
\affiliation{Paul Scherrer Institute, CH-5232 Villigen PSI, Switzerland}

\author{Luis Fern{\'a}ndez Barqu{\'i}n}
\affiliation{Department CITIMAC, Facultad de Ciencias, Universidad de Cantabria, 39005 Santander, Spain}

\author{Andreas Michels}\email[Electronic address: ]{andreas.michels@uni.lu}
\affiliation{Department of Physics and Materials Science, University of Luxembourg, 162A~Avenue de la Faiencerie, 1511~Luxembourg, Grand Duchy of Luxembourg}


\begin{abstract}
We experimentally report a hitherto unseen angular anisotropy in the polarized small-angle neutron scattering (SANS) cross section of a magnetically strongly inhomogeneous material. Based on an analytical prediction using micromagnetic theory, the difference between the spin-up and spin-down SANS cross sections is expected to show a spin-disorder-induced anisotropy. The effect is particularly pronounced in inhomogeneous magnetic materials, such as nanoporous ferromagnets, magnetic nanocomposites, or steels, which exhibit large nanoscale jumps in the saturation magnetization at internal pore-matrix or particle-matrix interfaces. Analysis of the experimental neutron data constitutes a method for determining the exchange-stiffness constant. Our results are generic to the nuclear-magnetic interference terms contained in the polarized magnetic neutron scattering cross section and might also be of relevance to other neutron techniques.
\end{abstract}

\date{\today}

\maketitle

 
{\it Introduction.} Polarized neutron scattering is one of the most powerful techniques for investigating the structure and dynamics of condensed matter, in particular magnetic materials and superconductors~\cite{tapan2006,boothroydbook}. Based on the seminal papers by Bloch, Schwinger, and Halpern and Johnson~\cite{bloch1936,bloch1937,schwinger1937,halpern39}, the theory of polarized neutron scattering has been worked out in the early 1960's by Maleev and Blume~\cite{maleyev63,blume63}. Several classic experimental studies~\cite{shull51,moon69,rekveldt71,drabkin72,okorokov1978,mezei86,schaerpf1993} have demonstrated the basic principles and paved the way for todays three-dimensional cryogenic polarization-analysis device (CRYOPAD)~\cite{tasset89,brown1993,tasset99,okorokov2001}. With this technique it becomes possible to measure 16~correlation functions, which provide important information on the nuclear and magnetic structure of materials (see Refs.~\cite{williams,lovesey} for textbook expositions of polarized neutron scattering).

Compared to unpolarized neutrons, the scattering cross section for polarized neutrons contains additional contributions~\cite{maleyev2002}. These are the familiar interference terms between the nuclear (structural) and magnetic scattering amplitudes and a purely magnetic-magnetic interference term (the so-called chiral function). In this paper, we exclusively concentrate on the {\it nuclear-magnetic interference terms}. Based on an analytical prediction using the continuum theory of micromagnetics, more  specifically for the transversal magnetization Fourier component~\cite{michelsPRB2016}, it is the central aim to experimentally search for the existence of a corresponding angular anisotropy in the nuclear-magnetic interference terms. This research makes a fundamental contribution to the understanding of polarized magnetic neutron scattering, and it widens the analysis capabilities of the polarized SANS technique by providing a method for determining the exchange constant.

We refer to the Supplemental Material~\cite{supmat2025} for additional micromagnetic calculations supporting the experimental neutron data.

{\it Experimental.} The theoretical considerations (see below) require a polycrystalline magnetic material with strong nanoscale spatial variations in the saturation magnetization, i.e., $M_{\mathrm{s}} = M_{\mathrm{s}}(\mathbf{r})$. Therefore, for the neutron experiments, we used inert-gas condensed nanoporous Fe~\cite{elmas09} and a melt-spun nanocrystalline $\mathrm{Fe_{89}Zr_{7}B_{3}Cu}$ alloy (Nanoperm)~\cite{michels03epl,michels06prb,suzuki2020,mathiasiucrj2022,rai2024}. The microstructure of the Fe sample consists of a distribution of nanosized pores in an Fe matrix, whereas the Nanoperm sample has a two-phase microstructure consisting of Fe nanoparticles that are embedded in an amorphous magnetic matrix of different magnetization. Hence, these specimens are characterized by large jumps $\Delta M$ in the magnitude of the saturation magnetization at internal pore-matrix and particle-matrix interfaces, $\mu_0\Delta M \cong 2.15 \, \mathrm{T}$ for Fe and $\mu_0 \Delta M \cong 1.5 \, \mathrm{T}$ for Nanoperm. The sample thicknesses for the SANS measurements were $\sim$$500 \, \mathrm{\mu m}$ (Fe) and $\sim$$100 \, \mathrm{\mu m}$ (Nanoperm). Unpolarized SANS investigations of these two materials along with details regarding sample synthesis and microstructural and magnetic characterization can be found in Refs.~\cite{elmas09,michels03epl,michels06prb,suzuki2020,mathiasiucrj2022,rai2024}.

\begin{figure}[tb!]
\centering
\resizebox{1.0\columnwidth}{!}{\includegraphics{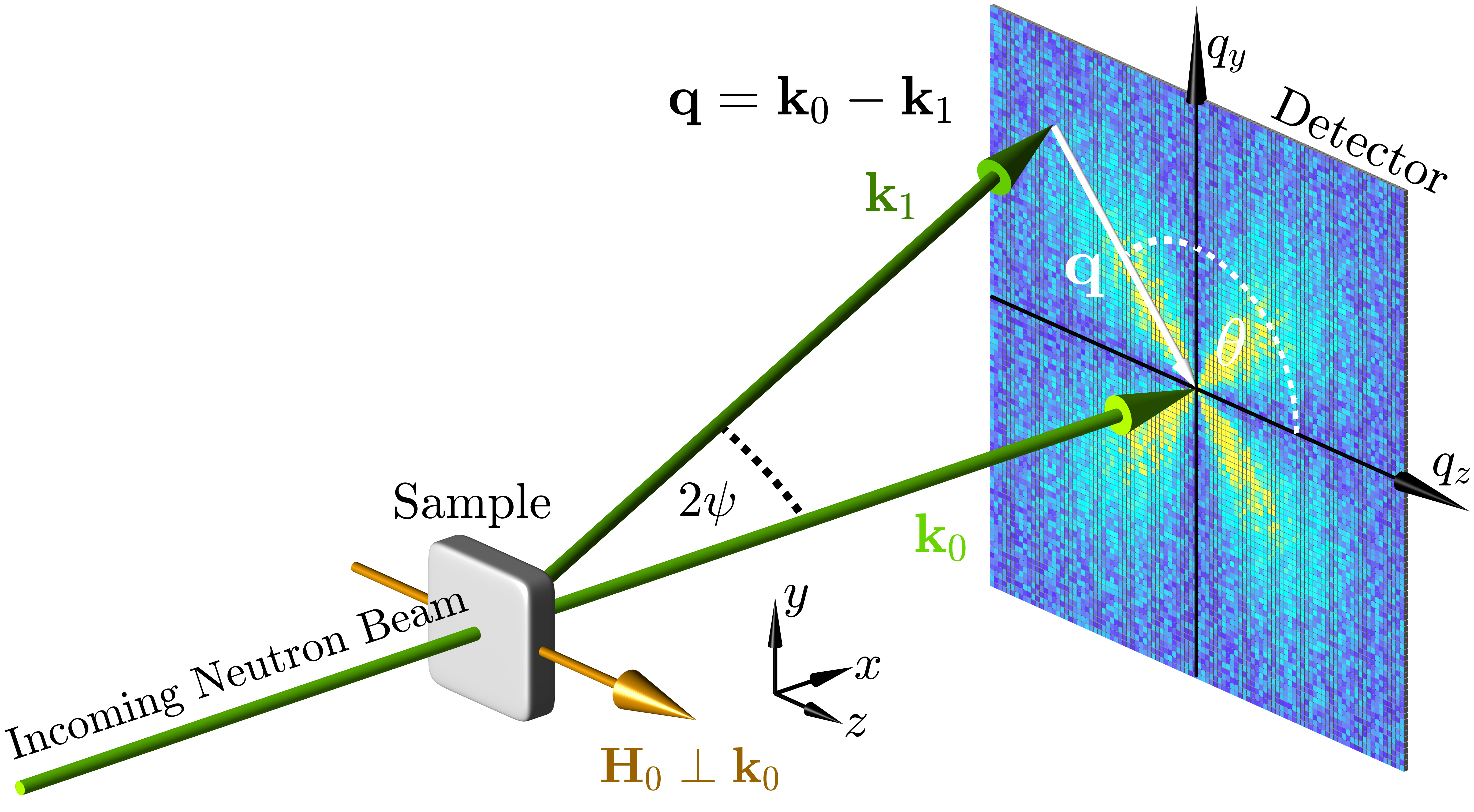}}
\caption{Sketch of the neutron scattering geometry. The neutron optical elements (polarizer and spin flipper) that are required to measure the two spin-resolved SANS cross sections are not drawn. The applied magnetic field $\mathbf{H}_0 \parallel \mathbf{e}_z$ is perpendicular to the wave vector $\mathbf{k}_0 \parallel \mathbf{e}_x$ of the incident neutron beam ($\mathbf{H}_0 \perp \mathbf{k}_0$). The momentum-transfer or scattering vector $\mathbf{q}$ is defined as the difference between $\mathbf{k}_0$ and $\mathbf{k}_1$, i.e., $\mathbf{q} = \mathbf{k}_0 - \mathbf{k}_1$. The angle $\theta = \angle(\mathbf{q}, \mathbf{H}_0)$ is used to describe the angular anisotropy of the recorded scattering pattern on the two-dimensional detector.}
\label{fig1}
\end{figure}

The neutron experiment was conducted at the instrument D33 at the Institut Laue-Langevin, Grenoble, France~\cite{dewhurst2016,illdoi-angani}. We used an incident polarized neutron beam with a mean wavelength of $\lambda = 4.65$~\AA \, and a wavelength broadening of $\Delta\lambda / \lambda = 10 \, \%$ (full width at half maximum). Two sample-to-detector distances ($13 \, \mathrm{m}$ and $5 \, \mathrm{m}$) allowed us to cover a $q$~range of $0.04 \, \mathrm{nm}^{-1} \lesssim q \lesssim 1.0 \, \mathrm{nm}^{-1}$. The external magnetic field $\mathbf{H}_0$ was provided by a superconducting magnet ($\mu_0 H_0^{\mathrm{max}} = 3 \, \mathrm{T}$) and applied perpendicular to the wave vector $\mathbf{k}_0$ of the incident neutron beam; see Fig.~\ref{fig1} for a schematic drawing of the experimental neutron setup. The beam was polarized by a magnetized FeSi multilayer mirror ($m = 3.6$), and an adiabatic resonance radio frequency (rf) spin flipper allowed us to reverse the initial neutron polarization. The flipping efficiency of the rf flipper was $\epsilon = 96 \, \%$, and the polarization of the beam was $P = 98 \, \%$ at $\lambda = 4.65$~\AA. Further neutron experiments under similar conditions have been performed at the ZOOM beamline~\cite{isisdoi-angani} at the ISIS Neutron and Muon Facility (Rutherford Appleton Laboratory, Didcot, United Kingdom) and at SANS-1 at the Paul Scherrer Institute (Villigen PSI, Switzerland). For SANS data reduction (correction for background scattering and polarization-dependent transmission), the GRASP software package was used~\cite{dewhurst2023}.

{\it Polarized SANS cross section.} 
The focus in our study lies on the difference $\Delta\Sigma = d \Sigma^{-} / d \Omega - d \Sigma^{+} / d \Omega$ between the flipper-off (``$-$'') and flipper-on (``$+$'') SANS cross sections. Neglecting nuclear spin-dependent scattering and the chiral function, which is expected to average out for statistically-isotropic polycrystalline magnetic materials, we can express $\Delta\Sigma$ as ($\mathbf{H}_0 \perp \mathbf{k}_0$, see Fig.~\ref{fig1})~\cite{michelsbook}:
\begin{eqnarray}
\label{sanspolperpdiff}
\Delta\Sigma = K \left[ (\widetilde{N} \widetilde{M}_z^{\ast} + \widetilde{N}^{\ast} \widetilde{M}_z) \sin^2\theta \right. 
\\ \left. - (\widetilde{N} \widetilde{M}_y^{\ast} + \widetilde{N}^{\ast} \widetilde{M}_y) \sin\theta \cos\theta \right] \nonumber ,
\end{eqnarray}
where $K = \frac{16 \pi^3}{V} b_{\mathrm{H}}$, $V$ is the scattering volume, $b_{\mathrm{H}} = 2.91 \times 10^8 \, \mathrm{A}^{-1}\mathrm{m}^{-1}$ is the magnetic scattering length in the small-angle regime (the atomic magnetic form factor is approximated by $1$, since we are dealing with forward scattering), $\widetilde{N}(\mathbf{q})$ and $\widetilde{\mathbf{M}}(\mathbf{q}) = \{ \widetilde{M}_x, \widetilde{M}_y, \widetilde{M}_z \}$ denote, respectively, the Fourier transforms of the nuclear scattering-length density and of the magnetization vector field $\mathbf{M}(\mathbf{r}) = \{ M_x, M_y, M_z \}$, $\theta$ is the angle between $\mathbf{H}_0 = H_0 \mathbf{e}_z$ and $\mathbf{q}$, so that $\mathbf{q} \cong q \{ 0, \sin\theta, \cos\theta \}$ in small-angle approximation, and the asterisks ``$*$'' mark the complex-conjugated quantity.

Equation~(\ref{sanspolperpdiff}) shows that there are two nuclear-magnetic interference terms contributing to $\Delta\Sigma$ (in the $\mathbf{H}_0 \perp \mathbf{k}_0$ geometry):~For isotropic $\widetilde{N}$ and $\widetilde{M}_z$, the first term exhibits the well-known $\sin^2\theta$ anisotropy, which has been observed countless times in polarized SANS experiments. It is the central aim of this paper to report on the experimental first-time observation of the second scattering term in Eq.~(\ref{sanspolperpdiff}), which allows the direct measurement of the exchange-stiffness constant. As we will detail in the following, this is accomplished in strongly inhomogeneous (regarding the spatial variation of the saturation magnetization) nanoporous Fe~\cite{elmas09} and in the two-phase nanocrystalline alloy Nanoperm~\cite{michels06prb}.

\begin{figure*}[ht!]
\centering
\resizebox{1.90\columnwidth}{!}{\includegraphics{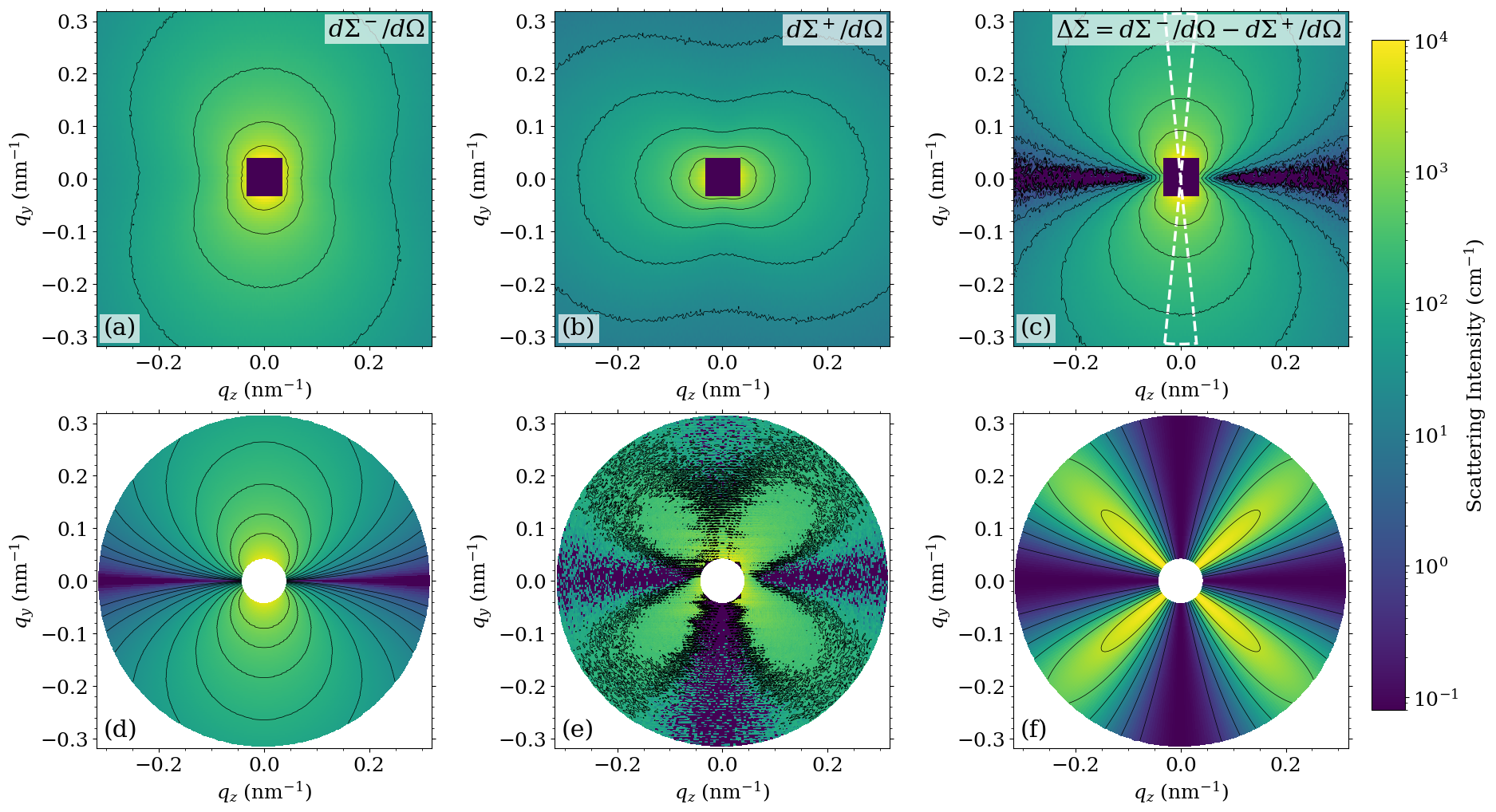}}
\caption{Neutron data analysis procedure. (a)~$d \Sigma^{-} / d \Omega$, (b)~$d \Sigma^{+} / d \Omega$, and (c)~$\Delta\Sigma = d \Sigma^{-} / d \Omega - d \Sigma^{+} / d \Omega$ of inert-gas condensed (igc) nanoporous Fe ($\mu_0 H_0 = 3.0 \, \mathrm{T}$). (d)~$\Delta\Sigma_{\theta = 90^{\circ}} \sin^2\theta$ extrapolated on the 2D detector using the $\Delta\Sigma$~data along the vertical direction from (c) (indicated by the white dashed lines). The data set shown in (d) corresponds to the first term in Eq.~(\ref{sanspolperpdifffinal}). (e)~Difference between the experimental data (c) and the extrapolated contribution~(d). Note that subfigures (a)$-$(e) show experimental data, while (f) features the analytical micromagnetic result for $\Delta\Sigma_{\mathrm{H}}$ [Eq.~(\ref{sanspolperpdiffffinal})]. Materials parameters for Fe were used~\cite{supmat2025}.}
\label{fig2}
\end{figure*}

{\it Micromagnetic SANS theory.} Theory predicts that in the two-dimensional $y$-$z$~detector plane (see Fig.~\ref{fig1}) the transversal magnetization Fourier component $\widetilde{M}_y = \widetilde{M}_y(q_x = 0, q_y, q_z)$ takes on the following form~\cite{michelsbook}:
\begin{equation}
\label{solmyqx0}
\widetilde{M}_y = \frac{p \left( \widetilde{H}_{\mathrm{p} y} - \widetilde{M}_z \sin\theta \cos\theta \right)}{1 + p \sin^2\theta} ,
\end{equation}
where $\widetilde{H}_{\mathrm{p} y}$ denotes the Cartesian component of the Fourier transform of the magnetic anisotropy field, $\widetilde{M}_z$ is the longitudinal magnetization Fourier component, and $p(q, H_0) = M_0/[H_0 (1 + l_{\mathrm{H}}^2 q^2)]$ is a known dimensionless function of $q$ and $H_0$, where $l_{\mathrm{H}}(H_0) = [2 A / (\mu_0 M_0 H_0)]^{1/2}$ denotes the micromagnetic exchange length of the field; $A$ is the exchange-stiffness constant, and $M_0 = \langle M_{\mathrm{s}}(\mathbf{r}) \rangle$ denotes the macroscopic saturation magnetization of the sample, which corresponds to the spatial average ($\langle ... \rangle$) of $M_{\mathrm{s}}(\mathbf{r})$. Equation~(\ref{solmyqx0}) results from the micromagnetic theory of the magnetic SANS cross section of bulk ferromagnets, which takes into account the isotropic exchange interaction, magnetic anisotropy, the magnetodipolar interaction as well as the external magnetic field~\cite{mysolution}.

If we assume that the nuclear scattering amplitude is isotropic, $\widetilde{N} = \widetilde{N}(q)$, and that $\widetilde{H}_{\mathrm{p} y}$ varies randomly in the plane perpendicular to $\mathbf{H}_0 \parallel \mathbf{e}_z$ (equal number of ``up'' and ``down'' orientations of $\widetilde{H}_{\mathrm{p} y}$ in a statistically-isotropic sample), then the corresponding averages over the direction of the anisotropy field vanish~\cite{angularanisotropy}. The $\widetilde{N} \widetilde{M}_y$ scattering contribution in Eq.~(\ref{sanspolperpdiff}) is then given by:
\begin{equation}
\label{nmycrosstermfinal}
2 \widetilde{N} \widetilde{M}_y \sin\theta \cos\theta = - \frac{2 p \widetilde{N} \widetilde{M}_z \sin^2\theta \cos^2\theta}{1 + p \sin^2\theta} ,
\end{equation}
where we have further assumed that $\widetilde{N}$, $\widetilde{M}_y$, and $\widetilde{M}_z$ are real-valued functions. Note the dominant angular $\sin^2\theta \cos^2\theta$~anisotropy of this term. Since $\widetilde{M}_z(q)$ represents, in the approach-to-saturation regime, the Fourier transform of the saturation magnetization profile $M_{\mathrm{s}}(r)$ of the sample, it is directly seen that the $\sin^2\theta \cos^2\theta$ contribution is expected to be observed for strongly inhomogeneous materials such as magnetic nanocomposites or porous ferromagnets [on top of the $\sin^2\theta$~anisotropy, compare Eq.~(\ref{sanspolperpdiff})]. On the other hand, when $M_{\mathrm{s}} = \mathrm{constant}$ throughout the sample, as is appropriate for homogeneous single-phase magnets, the corresponding scattering only shows up at the origin of reciprocal space and cannot be observed. We emphasize that both $\widetilde{M}_y$ and $\widetilde{M}_z$ depend on the applied field $H_0$, but that $\widetilde{M}_y$ tends to zero as $H_0 \rightarrow \infty$, while $\widetilde{M}_z$ takes on its maximum value ($\widetilde{M}_{\mathrm{s}}$) in this limit. Therefore, in addition to their different angular anisotropies, field-dependent experiments are key to unraveling the two contributions to Eq.~(\ref{sanspolperpdiff}).

Inserting Eq.~(\ref{nmycrosstermfinal}) into Eq.~(\ref{sanspolperpdiff}) we obtain for the difference cross section
\begin{eqnarray}
\label{sanspolperpdifffinal}
\Delta\Sigma = 2 K \widetilde{N} \widetilde{M}_z \sin^2\theta \left[ 1 + \frac{p \cos^2\theta}{1 + p \sin^2\theta} \right] .
\end{eqnarray}
Figure~1 in the Supplemental Material~\cite{supmat2025} displays the two-dimensional $\Delta\Sigma = \Delta\Sigma(q, \theta, H_0)$ [Eq.~(\ref{sanspolperpdifffinal})] at a series of applied magnetic fields. There, it is seen that overall a $\sin^2\theta$~type angular anisotropy prevails in the data at all fields and that with decreasing $H_0$ the pattern broadens.

Finally, subtracting the saturated term ($\propto \sin^2\theta$) in Eq.~(\ref{sanspolperpdifffinal}) yields the field-dependent contribution
\begin{eqnarray}
\label{sanspolperpdiffffinal}
\Delta\Sigma_{\mathrm{H}} = 2 K \widetilde{N} \widetilde{M}_z \frac{p \sin^2\theta \cos^2\theta}{1 + p \sin^2\theta} ,
\end{eqnarray}
which represents the main analytical result of this paper. Straightforward analysis shows that the maxima $\theta_{\mathrm{max}}$ of Eq.~(\ref{sanspolperpdiffffinal}) shift from about $45^{\circ}$ at high fields (small $p$) to about $30^{\circ}$ at low fields (large $p$)~\cite{supmat2025}. This provides a clear pathway towards identifying the angular anisotropy under question in experimental neutron data.

\begin{figure*}[tb!]
\centering
\resizebox{1.90\columnwidth}{!}{\includegraphics{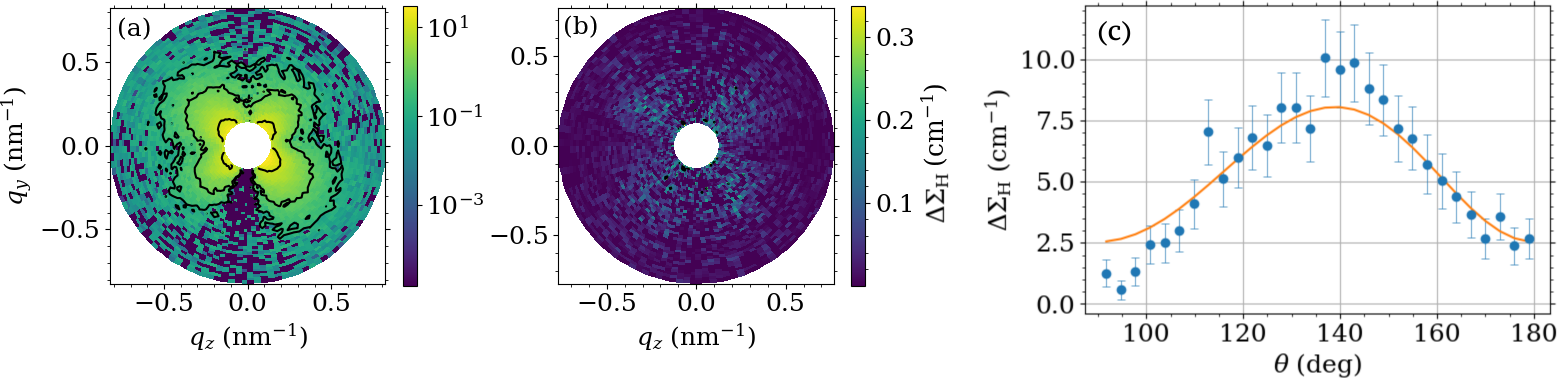}}
\caption{Experimental polarized SANS results of inert-gas condensed (igc) nanoporous Fe and nanocrystalline $\mathrm{Fe_{89}Zr_{7}B_{3}Cu}$ alloy (Nanoperm). (a)~$\Delta\Sigma_{\mathrm{H}}$ of igc Fe at $0.1 \, \mathrm{T}$; (b)~$\Delta\Sigma_{\mathrm{H}}$ of Nanoperm at $3.0 \, \mathrm{T}$; (c)~($\bullet$)~Experimental $\Delta\Sigma_{\mathrm{H}}(\theta)$ of igc Fe at $0.1 \, \mathrm{T}$, $q = 0.22 \, \mathrm{nm}^{-1}$, and for $90^{\circ} \lesssim \theta \lesssim 180^{\circ}$ [upper left quadrant in (a)]. Solid line:~Fit to Eq.~(\ref{sanspolperpdiffffinal}).}
\label{fig3}
\end{figure*}

{\it Experimental Results and Discussion.} The experimental data analysis procedure is explained in Fig.~\ref{fig2} and the polarized neutron results are summarized in Fig.~\ref{fig3}. For the case of inert-gas condensed nanoporous Fe at an applied magnetic field of $\mu_0 H_0 = 3 \, \mathrm{T}$, we show in Fig.~\ref{fig2}(a) and \ref{fig2}(b) the two half-polarized SANS cross sections $d \Sigma^{-} / d \Omega$ and $d \Sigma^{+} / d \Omega$, while the difference $\Delta\Sigma = d \Sigma^{-} / d \Omega - d \Sigma^{+} / d \Omega$ is displayed in Fig.~\ref{fig2}(c). At an angle of $\theta = 90^{\circ}$ [see the white dashed lines in Fig.~\ref{fig2}(c)], the second term in Eq.~(\ref{sanspolperpdifffinal}) vanishes and we obtain the ``usual'' nuclear-magnetic interference term $\Delta\Sigma_{\theta = 90^{\circ}} = 2 K \widetilde{N} \widetilde{M}_z$. This contribution depends on the magnitude $q$ of the scattering vector and under the assumption that both amplitudes $\widetilde{N}$ and $\widetilde{M}_z$ are isotropic (i.e., $\theta$~independent), we can generate (extrapolate) the corresponding 2D contribution $\Delta\Sigma_{\theta = 90^{\circ}} \sin^2\theta$ to Eq.~(\ref{sanspolperpdifffinal}) [Fig.~\ref{fig2}(d)]. These 2D data are then subtracted from the experimental $\Delta\Sigma$~data [Fig.~\ref{fig2}(c)] to approximately obtain the $\Delta\Sigma_{\mathrm{H}}$~term of Eq.~(\ref{sanspolperpdiffffinal}) [Figs.~\ref{fig2}(e) and \ref{fig2}(f)]. In this way we unravel the dominant $\sin^2\theta \cos^2\theta$~type angular anisotropy of interest.

The $\sin^2\theta \cos^2\theta$~type angular anisotropy is also observed in the nanoporous Fe sample at a lower field of $0.1 \, \mathrm{T}$ [Fig.~\ref{fig3}(a)] and also becomes visible in the polarized SANS data of the two-phase alloy Nanoperm at $3 \, \mathrm{T}$ [Fig.~\ref{fig3}(b)]. In Nanoperm, the anisotropy is less pronounced, which might be related to the fact that the jump in the saturation magnetization in this material is smaller than in the igc Fe sample. We also emphasize that for the field regime studied here both samples are within the approach-to-saturation regime~\cite{supmat2025}. Figure~\ref{fig3}(c) displays the angular variation of $\Delta\Sigma_{\mathrm{H}}$ of inert-gas condensed Fe at a field of $0.1 \, \mathrm{T}$ and for $q = 0.22 \, \mathrm{nm}^{-1}$. The solid line in Fig.~\ref{fig3}(c) represents a fit to Eq.~(\ref{sanspolperpdiffffinal}) with two free parameters:~the product $2 K \widetilde{N} \widetilde{M}_z$ is assumed to be constant at a fixed $q$~value ($2 K \widetilde{N} \widetilde{M}_z = 39.9 \pm 9.8 \, \mathrm{cm}^{-1}$) and the exchange-stiffness constant $A$ in the function $p$ is obtained as $A = (5.1 \pm 0.2) \times 10^{-11} \, \mathrm{J/m}$. The latter value fits well into the range of reported $A$~values for Fe~\cite{kronfahn03}.

{\it Conclusion.} We have theoretically predicted and experimentally verified the existence of a spin-disorder-induced angular anisotropy in the polarized magnetic small-angle neutron scattering cross section. In the approach-to-saturation regime, the result Eq.~(\ref{sanspolperpdiffffinal}) is of general relevance to magnetically inhomogeneous materials, such as nanoporous magnets, nanocomposites and permanent magnets, or steels, which exhibit a strong variation in the saturation magnetization $M_{\mathrm{s}}(\mathbf{r})$. Analysis of the angular dependence of $\Delta\Sigma_{\mathrm{H}}$ provides a means to determine the exchange constant. Since the nuclear-magnetic interference term under question ($\propto \widetilde{N} \widetilde{M}_y$) is generically contained in the polarized neutron scattering cross section it is of interest to verify its existence with other techniques such as neutron diffraction.

Michael P.\ Adams, Evelyn Pratami Sinaga, \v{S}tefan Li\v{s}\v{c}\'{a}k, Vladyslav M.\ Kuchkin, Andreas Haller, Thomas Schmidt, and Andreas Michels acknowledge financial support from the National Research Fund of Luxembourg (PRIDE MASSENA Grant, AFR Grant No.~15639149, AFR/23/17951349, and DeQuSky Grant No.~C22/MS/17415246). Vladyslav M.\ Kuchkin acknowledges financial support from the European Union's Horizon 2020 research and innovation program under the Marie Sk{\l}odowska-Curie grant agreement No.~101203692 (QUANTHOPF). Moreover, the authors thank the Science and Technology Facilities Council at ISIS, the Swiss spallation neutron source at the Paul Scherrer Institute, and the Institut Laue-Langevin for the provision of neutron beam time.


\begin{thebibliography}{38}
\bibitem [{ang()}]{angularanisotropy}%
  \BibitemOpen
  \href@noop {} {}\bibinfo {note} {A statistically-isotropic polycrystalline magnetic material may be characterized by random variations of the magnitude and direction of the magnetic anisotropy field, e.g., from one crystallite to another. In our experiment, a large external magnetic field is applied, which gives rise to a longitudinal ($z$) magnetization component that is close to the saturation value. For this scenario, the expectation values of the two transversal components of the anisotropy field vanish. Consequently, when averaging the $\widetilde{N} \widetilde{M}_y$~term in Eq.~(\ref{sanspolperpdiff}) [using Eq.~(\ref{solmyqx0})] over the directions of the anisotropy field, the $\widetilde{N} \widetilde{H}_{\mathrm{p} y}$~term vanishes, and only the averages over $\widetilde{N} \widetilde{M}_z$ remain.}\BibitemShut {Stop}%
\end{thebibliography}

%

\end{document}